\documentclass{IEEEtran}
\usepackage{amssymb}
\usepackage{amsfonts}
\usepackage{amsmath}
\usepackage{algorithm}
\usepackage{bm}
\usepackage{graphicx,subfigure,cite,multicol,multirow,diagbox,booktabs,array}
\usepackage{stfloats}

\usepackage[dvips]{color}
\usepackage{float}
\usepackage{subfig}%多个子图
\ifCLASSINFOpdf
\else
\fi
\begin{document}
\title{Enhanced RSS-based UAV Localization via Trajectory and Multi-base Stations }
\author{Yifan Li, Feng Shu, Baihua Shi, Xin Cheng, Yaoliang Song, and Jiangzhou Wang,~\emph{Fellow},~\emph{IEEE}
\thanks{Y. Li, B. Shi, X. Cheng, and Y. Song are with the School of Electronic and Optical Engineering, Nanjing University of Science and Technology, Nanjing 210094, China.}
\thanks{F. Shu is with the School of Information and Communication Engineering, Hainan University, Haikou 570228, China. and also with the School of Electronic and Optical Engineering, Nanjing University of Science and Technology, Nanjing 210094, China. Email: {shufeng0101@163.com}.}
\thanks{J. Wang is with the School of Engineering and Digital Arts, University of Kent, Canterbury CT2 7NT, U.K. (e-mail: j.z.wang@kent.ac.uk).}
}\maketitle

\begin{abstract}
To improve the localization precision of unmanned aerial vehicle (UAV), a novel  framework is established by jointly utilizing multiple measurements of received signal strength (RSS) from  multiple base stations (BSs) and multiple points on trajectory.  First,  a joint maximum likelihood (ML) of exploiting both trajectory  information and multi-BSs is proposed. To reduce its high complexity, two low-complexity localization methods are designed.  The first method is  from BS to  trajectory (BST), called LCSL-BST. First, fixing the $n$th BS, by exploiting multiple measurements along trajectory, the position of UAV  is computed by ML rule. Finally, all computed positions of UAV for different BSs are combined to form the resulting position. The second method reverses the order,  called LCSL-TBS.  We also derive the Cramer-Rao lower boundary (CRLB) of the joint ML method. From simulation results, we can see that the proposed joint ML and separate LCSL-BST methods have made a significant improvement over conventional ML method without use of trajectory knowledge in terms of location performance. The former achieves the joint CRLB and the latter is of low-complexity.
\end{abstract}
\begin{IEEEkeywords}
unmanned aerial vehicle (UAV), received signal strength (RSS), trajectory localization, maximum likelihood (ML), Cramer-Rao lower boundary (CRLB)
\end{IEEEkeywords}
\section{Introduction}
In the recent years, unmanned aerial vehicle (UAV) has been a hot topic, and technologies related to it are rapidly developed. Since the advantages in terms of three-dimensional (3D) mobility and flexible deployment, UAVs have been used in civilian and military areas, such as taking aerial photographs or videos, detecting enemy situations and destroying specific targets~\cite{8811579}~\cite{7470933}. Although currently there are not many actual cases of large-scale use of UAVs in the wireless communications, according to existing studies, UAVs may be involved in communication systems as mobile BSs or relays in the future~\cite{8125164}. Thus, when UAVs appear above our city, a critical question we need to consider is how to locate them quickly and  precisely. The position information of UAV is very helpful for us to plan the flight trajectory of UAV, so that we can cover a larger communication range with fewer UAVs.

In order to obtain the position information of UAV accurately, the appropriate localization technology should be selected. Over the past few decades, global positioning system (GPS) has been the preferred technology in positioning or navigation because of its reliability and accuracy in outdoor open situations. While in urban environments where buildings and people are crowded, or in battlefields, GPS signals are often weak or even blocked~\cite{9020301}. In these situations, we should consider the use of signals transmitted by the base stations (BSs) to determine the position of target. So we have turned our attention to other more practical geolocation techniques, including received signal strength (RSS), angle of arrival (AOA), time of arrival (TOA), and time difference of arrival (TDOA)~\cite{7776822}~\cite{8207615}~\cite{8290952}. This paper mainly focuses on RSS, because of the following advantages:  only a single antenna being required, simpler receivers, lower in cost, and low power conssumption~\cite{6735600}. Well-known algorithms for RSS geolocation include the Min-Max, Multiateration, Maximum Likelihood, and Least Squares methods.

Different from the traditional RSS-based 2D positioning model~\cite{2003On}, we propose a 3D UAV self-positioning system model where a moving UAV combines trajectory information and RSS measurements from surrounding base stations to locate itself. The paths between the UAV and BSs are under line-of-sight (LoS) conditions. Our main contributions  are summarized as follows:
\begin{enumerate}
\item To improve the localization accuracy, with the help of trajectory knowledge, a joint UAV ML localization method is proposed in multi-BS scenario. Compared to the case without use of trajectory knowledge, the proposed joint method makes a significant improvement in positioning accuracy. Then, the corresponding Cramer Rao lower bound (CRLB) is derived. Simulation results show that, the proposed method can achieve the joint CRLB.
\item To reduce the high computational complexity of the proposed joint  localization method above,  two low-complexity separate localization (LCSL) methods  are proposed. The first method is  from BS to  trajectory (BST), called LCSL-BST. First, fixing the $n$th BS, by exploiting multiple measurements along trajectory, the position of UAV  is computed by ML rule. Finally, all computed positions of UAV for different BSs are combined to form the resulting position. Similarly, the second method is from trajectory to BS (TBS), called LCSL-TBS. Compared with the joint UAV ML localization, the computational complexities of the two methods are reduced significantly. For large values of measurement standard deviation $\sigma$, the LCSL-BST makes a significant improvement over the ML  without using trajectory knowledge.
\end{enumerate}

\emph{\rm{\textbf{Notation}}:}  Matrices, vectors, and scalars are denoted by letters of bold upper case, bold lower case, and lower case, respectively. Signs $(\cdot)^T$, $\mid\cdot\mid$ and $\parallel\cdot\parallel$ represent transpose, modulus and norm. $\mathbf{J}_{l\times m}$ represents a $l\times m$ matrix with all elements equal to 1 and $K\times K$ identity matrix is denoted by $\mathbf{I}_K$. The expectation operator is denoted by $E[\cdot]$.
\begin{figure}[htb]
  \centering
  % Requires \usepackage{graphicx}s_Mod
  \includegraphics[width=0.45\textwidth]{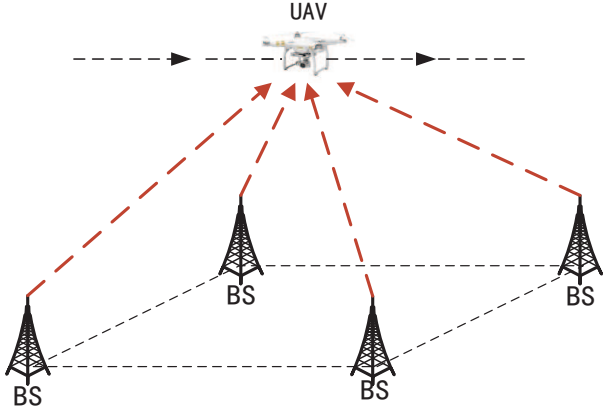}\\
  \caption{The UAV moves along the trajectory and receives signals from surrounding base stations.}\label{system model}
\end{figure}
\section{System Model}
In Fig.\ref{system model}, a system model, with an unmanned aerial vehicle (UAV) moving along a trajectory and receiving signals from the nearest $N$ BSs, is presented. Assuming that the BSs are located at $\mathbf{s}_n=[x_n^*,y_n^*,z_n^*],n=1,2,\cdots,N$, and the position of the UAV at time $k$ is $\mathbf{u}_k=[x_k,y_k,z_k], k=1,2,\cdots,K$. Suppose the time interval between any two points on the UAV trajectory is $\Delta t_i$, $i=1,2,\cdots,K-1$, and the UAV flies at a velocity of $\mathbf{v}_i=[v_{x,i},v_{y,i},v_{z,i}]$ at $\Delta t_i$. Hence, coordinate of the UAV at time $k$ can be rewritten as
\begin{equation}
%\begin{aligned}
\mathbf{u}_k=\mathbf{u}_1+\sum_{i=1}^{k-1}\mathbf{v}_i\Delta t_i=\mathbf{u}_1+\Delta\mathbf{u}_k,\label{u_k}
%\end{aligned}
\end{equation}
where
\begin{equation}
\begin{aligned}
\Delta\mathbf{u}_k&=[\Delta x_k,\Delta y_k,\Delta z_k]\\
&=\bigg[\sum_{i=1}^{k-1}v_{x,i}\Delta t_i,\sum_{i=1}^{k-1}v_{y,i}\Delta t_i,\sum_{i=1}^{k-1}v_{z,i}\Delta t_i\bigg],\label{delta_u_k}
\end{aligned}
\end{equation}
where $\Delta\mathbf{u}_1=[0,0,0]$. Using the log-distance path loss propagation model~\cite{2012Modeling}~\cite{2020Bayesian}, the RSS measurements coming from the $n$th BS and received by the UAV at time $k$, is given by
\begin{equation}
\begin{aligned}
r_{k,n}&=\alpha_n+a_{k,n}+w_{k,n}\label{rss_single},\\
a_{k,n}&=10\gamma\log_{10}\left(\frac{d_0}{d_{k,n}}\right),
\end{aligned}
\end{equation}
where $d_0$ is a predefined reference distance. $\alpha_n$ denotes the reference power in dBm at $d_0$ of the $n$th BS, and is assumed to be unknown. $\gamma$ stands for the path loss exponent, and its values generally range between $\gamma=2$ and $\gamma=5$~\cite{6030635}. $w_{k,n}$ is a zero-mean Gaussian random variable with known standard deviation $\sigma_{k,n}>0$. The Euclidean distance between the $n$th BS and the UAV at time $k$ can be verified as
\begin{equation}
\begin{aligned}
d_{k,n}=&\big[(x_k-x_n^*)^2+(y_k-y_n^*)^2+(z_k-z_n^*)^2\big]^{\frac{1}{2}}\\
=&\big[(x_1+\Delta x_k-x_n^*)^2+(y_1+\Delta y_k-y_n^*)^2\\
&+(z_1+\Delta z_k-z_n^*)^2\big]^{\frac{1}{2}}\\
=&\big[(x_1-(x_n^*-\Delta x_k))^2+(y_1-(y_n^*-\Delta y_k))^2\\
&+(z_1-(z_n^*-\Delta z_k))^2\big]^{\frac{1}{2}}.\label{distance}
\end{aligned}
\end{equation}
From another perspective, the system model can be viewed as the UAV fixed at $\mathbf{u}_1$, while the $n$th BS moves from the initial position $\mathbf{s}_n$ to the $\mathbf{s}_{k,n}=[x_n^*-\Delta x_k,y_n^*-\Delta y_k,z_n^*-\Delta z_k]$ at time $k$. Obviously, the BS doesn't move in reality, so they are called virtual moving BSs in our assumption.

Collecting all measurement values from $N$ virtual moving BSs at all $K$ times forms the following ${K\times N}$ RSS  matrix
\begin{equation}
\mathbf{R}=\bm{\alpha}\otimes\mathbf{J}_{K\times1}+\mathbf{A}+\mathbf{W},\label{rss}
\end{equation}
where
\begin{equation}
\begin{aligned}
%\begin{split}
&\mathbf{R}=\begin{bmatrix}
r_{1,1}&\dots&r_{1,N}\\
\vdots&\ddots&\vdots\\
r_{K,1}&\dots&r_{K,N}
\end{bmatrix}
\mathbf{A}=\begin{bmatrix}
a_{1,1}&\dots&a_{1,N}\\
\vdots&\ddots&\vdots\\
a_{K,1}&\dots&a_{K,N}
\end{bmatrix}\\
&\qquad\qquad\quad\mathbf{W}=\begin{bmatrix}
w_{1,1}&\dots&w_{1,N}\\
\vdots&\ddots&\vdots\\
w_{K,1}&\dots&w_{K,N}
\end{bmatrix},
\end{aligned}
\end{equation}
and $\bm{\alpha}=[\alpha_1,\alpha_2,\cdots,\alpha_N]$, $\otimes$ denotes the Kronecker product.

In order to obtain the maximum likelihood (ML) estimator of parameters, using the vec operator to transform the matrices to vectors  yields  the following form
\begin{equation}
\mathbf{r}={\rm vec}\left(\bm{\alpha}\otimes\mathbf{J}_{K\times 1}\right)+\mathbf{a}+\mathbf{w},
\end{equation}
where $\mathbf{r}={\rm vec}(\mathbf{R})$, $\mathbf{a}={\rm vec}(\mathbf{A})$ and $\mathbf{w}={\rm vec}(\mathbf{W})$. Based on the basic theorems of the variance,
\begin{equation}
E(w_{k_1,n_1}w_{k_2,n_2})=
\begin{cases}\sigma_{k_1,n_1}^2,\quad&k_1=k_2\quad and\quad n_1=n_2\\
0,\quad&k_1\neq k_2\quad or\quad n_1\neq n_2
\end{cases}
\end{equation}
where $1\le k_1,k_2\le K$ and $1\le n_1,n_2\le N$, the covariance matrix of $\mathbf{w}$ is given by
\begin{equation}
\begin{aligned}
\mathbf{C}&=E[(\mathbf{w}-E[\mathbf{w}])(\mathbf{w}-E[\mathbf{w}])^T]\\
&={\rm diag}\{\mathbf{V}_1,\mathbf{V}_2,\dots,\mathbf{V}_K\},
\end{aligned}
\end{equation}
where $\mathbf{V}_k={\rm diag}\{\sigma_{k,1}^2,\sigma_{k,2}^2,\cdots,\sigma_{k,N}^2\}$.
\section{Proposed Localization Methods with the Aid of Trajectory}
In this section, to achieve a high-precision localization,  three methods are proposed: an optimal joint ML  method and two low-complexity separate methods.  The latter two  methods make a position estimate along   BS dimension and the trajectory dimension separately to achieve a low-complexity at the expense of some performance loss.
\subsection{Proposed Joint Trajectory-aided ML Localization Method}
According to the system model, the localization of the UAV position $\mathbf{u}_1$ is based on RSS measurements from $N$ BSs at all $K$ times and assuming that the reference power of each BS is equal to $\alpha$. The probability density function (pdf) of RSS measurements $\mathbf{r}$ given $\mathbf{u}_1$ and $\alpha$, is denoted by %$f(\mathbf{r}|\mathbf{u},\alpha)$
\begin{equation}
\begin{aligned}
&f(\mathbf{r}|\mathbf{u}_1,\alpha)=M\exp\bigg\{-\frac{1}{2}(\mathbf{r}-\mathbf{a}-\alpha\mathbf{J}_{NK\times1})^T\mathbf{C}^{-1}\cdot\\
&\qquad\qquad\qquad\qquad\quad(\mathbf{r}-\mathbf{a}-\alpha\mathbf{J}_{NK\times1})\bigg\}\\
&=M\exp\bigg\{-\frac{1}{2}\sum_{k=1}^K\sum_{n=1}^N\sigma_{k,n}^{-2}(r_{k,n}-a_{k,n}-\alpha)^2\bigg\},\label{joint}
\end{aligned}
\end{equation}
%\frac{1}{(2\pi)^{NK/2}\vert\mathbf{C}\vert^{1/2}}
where $M=\left((2\pi)^{NK}\vert\mathbf{C}\vert\right)^{-\frac{1}{2}}$. Given measurement data $\mathbf{r}$, the above function is the likelihood function of two parameters $\mathbf{u}_1$ and $\alpha$.  Maximizing the likelihood function with respect to $\mathbf{u}_1$ and $\alpha$ is
equivalent to  minimizing the function
\begin{equation}
Q(\mathbf{u}_1,\alpha)=\sum_{k=1}^K\sum_{n=1}^N\sigma_{k,n}^{-2}(r_{k,n}-a_{k,n}-\alpha)^2.\label{Q}
\end{equation}
%(\mathbf{r}-\alpha\mathbf{J}_{NK\times1}-\mathbf{a})^T\mathbf{C}^{-1}(\mathbf{r}-\alpha\mathbf{J}_{NK\times1}-\mathbf{a})

Fixing parameter $\mathbf{u}_1$, the optimal value  $\hat{\alpha}$ of minimizing the above function is given by
\begin{equation}
\begin{aligned}
{\hat{\alpha}}=\frac{1}{NK}\sum_{k=1}^K\sum_{n=1}^N(r_{k,n}-a_{k,n}).\label{alpha}
\end{aligned}
\end{equation}
Substituting (\ref{alpha}) back in (\ref{Q}) yields
\begin{equation}
\begin{aligned}
Q(\mathbf{u}_1)=\sum_{k=1}^K\sum_{n=1}^N\sigma_{k,n}^{-2}&\bigg(r_{k,n}-a_{k,n}-\\
&\frac{1}{NK}\sum_{k=1}^K\sum_{n=1}^N(r_{k,n}-a_{k,n})\bigg)^2.\label{q_u1}
\end{aligned}
\end{equation}
%\begin{equation}
%\begin{aligned}
%&Q(\mathbf{u}_1)=[\mathbf{B}(\mathbf{r}-\mathbf{a})]^T\mathbf{C}^{-1}[\mathbf{B}(\mathbf{r}-\mathbf{a})]\\
%&\mathbf{B}=\mathbf{I}_{NK}-(NK)^{-1}\mathbf{J}_{NK\times1}\mathbf{J}_{NK\times1}^T.
%\end{aligned}
%\end{equation}
Maximizing $Q(\mathbf{u}_1)$ to obtain the optimal $\hat{\mathbf{u}}_1$ is done by performing the following nonlinear optimization
\begin{equation}
\hat{\mathbf{u}}_1={\rm{argmin}}\left\{Q(\mathbf{u}_1)\right\},\label{ML estimator}
\end{equation}
where $\hat{\mathbf{u}}_1=[\hat{x}_1,\hat{y}_1,\hat{z}_1]^T$. For this nonlinear optimization problem, the grid search is a reliable solution. The computational complexity of each search is $4(KN)^2$ real multiplications.
\subsection{Proposed LCSL-BST}
Suppose the UAV only receives signals from one BS at all $K$ times. So if we want to locate the UAV's position according to all  $N$ BSs, each BS is virtually viewed as moving along the trajectory separately. Finally, the result is a weighted combination of these $N$ estimations. Assuming the RSS measurement $\mathbf{r}_n$ from the $n$th BS is
\begin{equation}
\mathbf{r}_n=\alpha\mathbf{J}_{K\times1}+\mathbf{a}_n+\mathbf{w}_n,
\end{equation}
where $\mathbf{r}_n=[r_{1,n},\cdots,r_{K,n}]^T$ , $\mathbf{a}_n=[a_{1,n},\cdots,a_{K,n}]^T$, and $\mathbf{w}_n=[w_{1,n},\cdots,w_{K,n}]^T$. The objective function (\ref{q_u1}) is rewritten as
\begin{equation}
Q_n(\mathbf{u}_1)=\sum_{k=1}^K\sigma_{k,n}^{-2}\bigg(r_{k,n}-a_{k,n}-\frac{1}{K}\sum_{k=1}^K(r_{k,n}-a_{k,n})\bigg)^2,
\end{equation}
and then perform the optimization based on the RSS measurement from BS $n$ at all $K$ times
\begin{equation}
\tilde{\mathbf{u}}_{1,n}={\rm{argmin}}\left\{Q_n(\mathbf{u}_1)\right\}.
\end{equation}
%\begin{equation}
%\tilde{\mathbf{u}}_{1,n}={\rm{argmin}}\left\{Q_n(\mathbf{u}_1)\right\}
%\end{equation}
%where $\tilde{\mathbf{u}}_{1,n}=[\tilde{x}_{1,n},\tilde{y}_{1,n},\tilde{z}_{1,n}]^T$.
%\begin{equation}
%\begin{aligned}
%&\mathbf{C}_n=\sigma_n^2\mathbf{I}_K\\
%&\mathbf{B}_n=\mathbf{I}_K-K^{-1}\mathbf{J}_{K\times1}\mathbf{J}_{K\times1}^T
%\end{aligned}
%\end{equation}

Since each estimate is based on RSS measurements from different BSs, the location result after the $n$th estimate, corresponding to the $n$th BS, is designed as the combination of the current estimate $\tilde{\mathbf{u}}_{1,n}$ and the location result $\hat{\mathbf{u}}_{1,n-1}$ after the previous estimate. Then we can get $\hat{\mathbf{u}}_{1,n}=f_n(\hat{\mathbf{u}}_{1,n-1},\tilde{\mathbf{u}}_{1,n})$, where function $f_n()$ denotes a linear combination and the weighting coefficients are assumed to be equal for each term. So, the estimation result associated with the $n$th BS is
\begin{equation}
\begin{aligned}
f_n(\hat{\mathbf{u}}_{1,n-1},\tilde{\mathbf{u}}_{1,n})=\frac{1}{n}\sum_{m=1}^n\tilde{\mathbf{u}}_{1,m}=\frac{n-1}{n}\hat{\mathbf{u}}_{1,n-1}+\frac{1}{n}\tilde{\mathbf{u}}_{1,n},
%&\tilde{\mathbf{u}}_{n-1}=\frac{1}{n-1}\sum_{t=1}^{n-1}\hat{\mathbf{u}}_t
%&=\frac{1}{n}\sum_{t=1}^{n-1}\hat{\mathbf{u}}_t+\frac{1}{n}\hat{\mathbf{u}}_n\\
%&=\frac{n-1}{n}\tilde{\mathbf{u}}_{n-1}+\frac{1}{n}\hat{\mathbf{u}}_n
\end{aligned}
\end{equation}
where $\hat{\mathbf{u}}_{1,n-1}=\frac{1}{n-1}\sum_{m=1}^{n-1}\tilde{\mathbf{u}}_{1,m}$ and the final positioning result is
\begin{equation}
\hat{\mathbf{u}}_1=f_N(\hat{\mathbf{u}}_{1,N-1},\tilde{\mathbf{u}}_{1,N})=\frac{N-1}{N}\hat{\mathbf{u}}_{1,N-1}+\frac{1}{N}\tilde{\mathbf{u}}_{1,N},
\end{equation}
where $\hat{\mathbf{u}}_1=[\hat{x}_1,\hat{y}_1,\hat{z}_1]^T$. The grid search steps for this method are similar to the previous one, with a complexity of $4K^2N$ real multiplications per search.
\subsection{Proposed LCSL-TBS}
Assuming that the UAV locates itself separately according to the RSS measurements from the virtual BSs at each time. The RSS measurements obtained at time $k$ is
\begin{equation}
\mathbf{r}_k=\alpha\mathbf{J}_{N\times1}+\mathbf{a}_k+\mathbf{w}_k,
\end{equation}
where $\mathbf{r}_k=[r_{k,1},\cdots,r_{k,N}]^T$ , $\mathbf{a}_k=[a_{k,1},\cdots,a_{k,N}]^T$ and $\mathbf{w}_k=[w_{k,1},\cdots,w_{k,N}]^T$.  The object function $Q$ at time $k$ is given by
\begin{equation}
Q_k(\mathbf{u}_1)=\sum_{n=1}^N\sigma_{k,n}^{-2}\bigg(r_{k,n}-a_{k,n}-\frac{1}{N}\sum_{n=1}^N(r_{k,n}-a_{k,n})\bigg)^2,
\end{equation}
%\begin{equation}
%\tilde{\mathbf{u}}_{1,k}=\rm{argmin}\left\{[\mathbf{B}_k(\mathbf{r}_k-\mathbf{a}_k)]^T\mathbf{C}_k^{-1}[\mathbf{B}_k(\mathbf{r}_k-\mathbf{a}_k)]\right\}
%\end{equation}
%where $\tilde{\mathbf{u}}_{1,k}=[\hat{x}_{1,k},\hat{y}_{1,k}]$ and
%\begin{equation}
%\begin{aligned}
%&\mathbf{C}_k=\mathbf{V}\\
%&\mathbf{B}_k=\mathbf{I}_N-N^{-1}\mathbf{J}_{N\times1}\mathbf{J}_{N\times1}^T
%\end{aligned}
%\end{equation}
then, the optimization performed at time $k$ is
\begin{equation}
\tilde{\mathbf{u}}_{1,k}={\rm{argmin}}\left\{Q_k(\mathbf{u}_1)\right\}.
\end{equation}
Similar to the previous section, combining $\tilde{\mathbf{u}}_{1,k}$ at all the  $K$ time generates the final result
\begin{equation}
\hat{\mathbf{u}}_1=\frac{1}{K}\sum_{k=1}^K\tilde{\mathbf{u}}_{1,k}=\frac{K-1}{K}\hat{\mathbf{u}}_{1,K-1}+\frac{1}{K}\tilde{\mathbf{u}}_{1,K},
\end{equation}
where $\hat{\mathbf{u}}_{1,K-1}=\frac{1}{K-1}\sum_{m=1}^{K-1}\tilde{\mathbf{u}}_{1,m}$. The computational complexity of each search is $4KN^2$ real multiplications.

%The computational complexity of each estimation of these three methods is as follows : $4(NK)^2$ times multiplication and $6(NK)^2$ times addition or subtraction for joint ML trajectory localization method. $4NK^2$ times multiplication and $6NK^2+2NK$ times addition or subtraction for separate method I. $4KN^2$ times multiplication and $6KN^2$ times addition or subtraction for separate method II.

\section{Simulation Results and Discussion}
To illustrate the concepts and algorithms discussed in this paper, we will show some numerical results of the proposed methods in this section. According to the system model and equation (\ref{ML estimator}), grid search is a reliable method to find the point $\hat{\mathbf{u}}_1$ that minimizes the objective function $Q$. While there is an obvious fact that 3D grid search has high complexity. In order to facilitate the simulation, the UAV is set to fly at a known altitude, thus turning the problem into a 2D grid search. In the simulations, the flight altitude of UAV is $z=100$m, $K=10$, $\mathbf{v}_1=\dots=\mathbf{v}_{K-1}=[10,0]$ and $\Delta t_1=\dots=\Delta t_{K-1}=5$s. For the BSs, $N=6$ and BSs are located on the corners of a hexagon which is centered at origin. The distance between two adjacent BSs is 1km and the height of each BS is 20 meters. The error terms $w_{k,n}$ in (\ref{rss_single}) are all simulated as independent zero-mean Gaussian random variables with identical variance $\sigma^2$. To determine the $\hat{\mathbf{u}}_1$, a 2km$\times$2km area of interest (AOI) is set up and the area is divided into a 10m$\times$10m grid. The miss distance is the distance between the $\hat{\mathbf{u}}_1$ and $\mathbf{u}_1$. The average miss distance is computed from 1000 miss distance simulation runs.
\begin{figure}[h]
  \centering
  % Requires \usepackage{graphicx}s_Mod
  \includegraphics[width=0.45\textwidth]{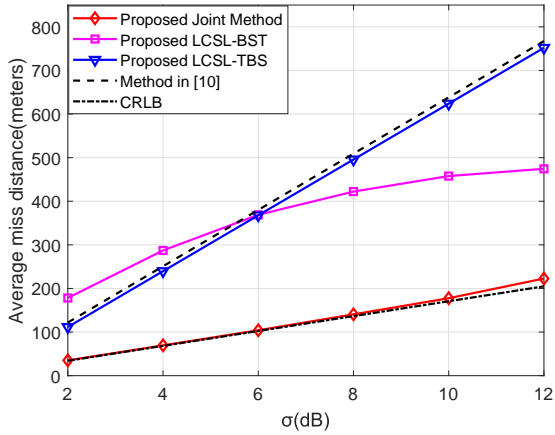}\\
  \caption{Average miss distance versus the measurement standard deviation $\sigma$.}\label{amd_sigma}
\end{figure}

Fig.\ref{amd_sigma} plots the average miss distance as a function of the measurement standard deviation $\sigma$, where $\gamma=3.3$.  The proposed joint method makes a significant improvement in the positioning accuracy over existing method \cite{2003On}. More importantly, the  proposed joint method can achieve the joint CRLB.  The proposed LCSL-TBS is slightly better than  existing method \cite{2003On} for almost all values of measurement standard deviation $\sigma$. The proposed LCSL-BST performs better than existing method \cite{2003On} for large values of measurement standard deviation $\sigma$, for example, $\sigma$ is larger than 6dB.
\begin{figure}[ht]
  \centering
  % Requires \usepackage{graphicx}s_Mod
  \includegraphics[width=0.45\textwidth]{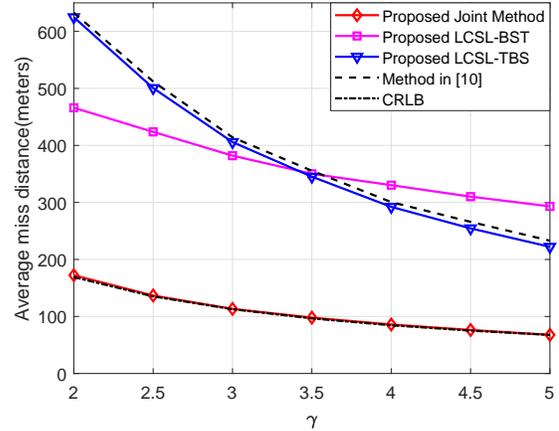}\\
  \caption{Average miss distance versus the path loss exponent $\gamma$.}\label{amd_gamma}
\end{figure}

Fig.\ref{amd_gamma} demonstrates the average miss distance as a function of the path loss exponent $\gamma$, where $\sigma=6$dB.  Observing this figure, we find the same performance tendency as Fig.\ref{amd_sigma}. In particular,  the proposed LCSL-BST performs better than existing method \cite{2003On} for small values of path loss exponent $\gamma$, for example, $\gamma$ is less than 3.5.
\begin{figure}[ht]
  \centering
  % Requires \usepackage{graphicx}s_Mod
  \includegraphics[width=0.45\textwidth]{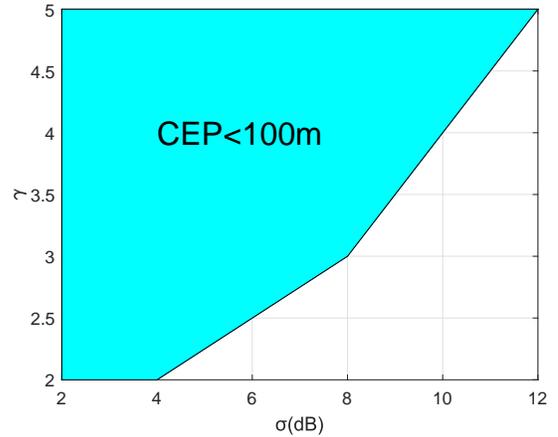}\\
  \caption{The measurement standard deviation $\sigma$ and the path loss exponent $\gamma$ satisfying the requirement $\rm{CEP}< 100\rm{m}$ based on the joint method.}\label{cep}
\end{figure}

Fig.\ref{cep} illustrates the circular error probability (CEP) of the joint localization method, the cyan
 area is the area of satisfying $\rm{CEP}< 100\rm{m}$ for the values of the measurement standard deviation $\sigma$ and path loss exponent $\gamma$. The CEP expression is given in~\cite{1984Statistical}.
%\begin{figure*}[t]  %[htbp]中的h是浮动的意思
%    \centering    %居中
%  %   \setcounter{figure}{4}
%
%    \subfigure[Average miss distance vs $\sigma$] % 第一张子图
%    {
%        \begin{minipage}[t]{0.48\textwidth}
%            \centering          %子图居中
%            \includegraphics[width=\textwidth]{sigma_6bs.eps}
%        \end{minipage}%
%    }
%    \subfigure[Average miss distance vs $\gamma$] % 第二张子图
%    {
%        \begin{minipage}[t]{0.48\textwidth}
%            \centering      %子图居中
%            \includegraphics[width=\textwidth]{gamma_6bs.eps}
%        \end{minipage}
%    }%
%
%    \caption{Average miss distance versus the measurement standard deviation $\sigma$ and the path loss exponent $\gamma$.} %  % 大图名称
% %   \label{fig13}  %图片引用标记
%\end{figure*}

\section{Conclusions}
In our work, an enhanced RSS-based UAV localization model has been proposed. This model jointly exploits the trajectory information and RSS measurements from multiple BSs to improve  the UAV localization performance. First, a joint ML localization method and two low-complexity separate methods were proposed in section III. The joint CRLB was also derived as a performance benchmark. The simulation results have shown that, compared with exsiting method without use of trajectory knowledge, the proposed joint method makes a significant improvement in positioning accuracy. The LCSL-BST and LCSL-TBS achieve a low-complexity at the expense of performance loss compared to the joint method.

\section*{Appendix:~Derivation of joint CRLB}
In order to obtain the Cramer-Rao lower bound (CRLB) on the location-error variance of the joint ML trajectory method, we extend the derivation in~\cite{2003On} to our work. Assuming $\mathbf{C}=\sigma^2\mathbf{I}_{NK}$, then
\begin{equation}
\rm{CRLB}=\sigma^2\mathbf{G}^{-1},
\end{equation}
%where
%\begin{equation}
%\mathbf{G}=\begin{bmatrix}
%g_{1,1}&\dots&g_{1,2K}\\
%\vdots&\ddots&\vdots\\
%g_{2K,1}&\dots&g_{2K,2K}
%\end{bmatrix}
%\end{equation}
where $\mathbf{G}$ is a $3\times 3$ symmetrical matrix and its elements can be expressed as
\begin{equation}
\begin{aligned}
g_{i,j}=&\mathbf{a}_{\mathbf{u}_1(i)}^T\mathbf{a}_{\mathbf{u}_1(j)}-\frac{1}{NK}\left(\mathbf{J}_{NK\times1}^T\mathbf{a}_{\mathbf{u}_1(i)})(\mathbf{J}_{NK\times1}^T\mathbf{a}_{\mathbf{u}_1(j)}\right)\label{g}\\
=&\mathbf{a}_{\mathbf{u}_1(i)}^T\mathbf{a}_{\mathbf{u}_1(j)}-\frac{1}{NK}\sum\mathbf{a}_{\mathbf{u}_1(i)}\sum\mathbf{a}_{\mathbf{u}_1(j)}\\
=&p_{i,j}-q_{i,j},
\end{aligned}
\end{equation}
where $\mathbf{u}_1=[x_1,y_1,z_1]$, $1\le i,j\le 3$, $\mathbf{u}_1(i)$ or $\mathbf{u}_1(j)$ represents the $i$th or $j$th element of the vector $\mathbf{u}_1$ and
\begin{equation}
\mathbf{a}_{\mathbf{u}_1(i)}=\frac{\partial\mathbf{a}}{\partial\mathbf{u}_1(i)}.
%\mathbf{a}_{\mathbf{u}_1(j)}=\frac{\partial\mathbf{a}}{\partial\mathbf{u}_1(j)}
\end{equation}
%\begin{equation}
%\mathbf{a}_{\bm\varphi(i)}=\frac{\partial\mathbf{a}}{\partial\bm\varphi(i)};\mathbf{a}_{\bm\varphi(j)}=\frac{\partial\mathbf{a}}{\partial\bm\varphi(j)}
%\end{equation}
%since $i$ and $j$ can be replaced by $2k-1$ or $2k$, $\mathbf{u}(2k-1)=x_k$, $\mathbf{u}(2k)=y_k$,
Then equations (\ref{rss_single}) and (\ref{distance}) lead to the following $NK\times1$ vectors
\begin{equation}
%\begin{aligned}
\mathbf{a}_{\mathbf{u}_1(i)}=\beta[\mathbf{a}_{\mathbf{u}_1(i),1},\mathbf{a}_{\mathbf{u}_1(i),2},\dots,\mathbf{a}_{\mathbf{u}_1(i),K}]^T,\label{a_qiudao}
%&\mathbf{a}_{y_k}=\beta\left[\overbrace{0,\dots,0}^{(k-1)N},\frac{y_k-y_1}{d_{k,1}^2},\dots,\frac{y_k-y_N}{d_{k,N}^2},\overbrace{0,\dots,0}^{(K-k)N}\right]^T
%\end{aligned}
\end{equation}
where $\beta=-(10\gamma)/\ln10$ and
\begin{equation}
\begin{aligned}
&\mathbf{a}_{\mathbf{u}_1(1),k}=\left[\frac{x_1+\Delta x_k-x_1^*}{d_{k,1}^2},\dots,\frac{x_1+\Delta x_k-x_N^*}{d_{k,N}^2}\right],\\
&\mathbf{a}_{\mathbf{u}_1(2),k}=\left[\frac{y_1+\Delta y_k-y_1^*}{d_{k,1}^2},\dots,\frac{y_1+\Delta y_k-y_N^*}{d_{k,N}^2}\right],\\
&\mathbf{a}_{\mathbf{u}_1(3),k}=\left[\frac{z_1+\Delta z_k-z_1^*}{d_{k,1}^2},\dots,\frac{z_1+\Delta z_k-z_N^*}{d_{k,N}^2}\right].
\end{aligned}
\end{equation}
Then we can get
%\begin{equation}
%\mathbf{a}_{\mathbf{u}_1(i)}^T\mathbf{a}_{\mathbf{u}_1(j)}=\sum_{k=1}^K\mathbf{a}_{\mathbf{u}_1(i),k}\mathbf{a}_{\mathbf{u}_1(j),k}^T
%\end{equation}
%and
\begin{equation}
\begin{cases}
p_{1,1}=\beta^2\sum_{k=1}^K\sum_{n=1}^N\frac{(x_1+\Delta x_k-x_n^*)^2}{d_{k,n}^4}\\
%&-\frac{\beta^2}{NK}\left(\sum_{k=1}^K\sum_{n=1}^N\frac{x_1+\Delta x_k-x_n^*}{d_{k,n}^2}\right)^2\\
%p_{2,2}=\beta^2\sum_{k=1}^K\sum_{n=1}^N\frac{(y_1+\Delta y_k-y_n^*)^2}{d_{k,n}^4}\\
p_{1,2}=\beta^2\sum_{k=1}^K\sum_{n=1}^N\frac{(x_1+\Delta x_k-x_n^*)(y_1+\Delta y_k-y_n^*)}{d_{k,n}^4}\\
%p_{1,3}=\beta^2\sum_{k=1}^K\sum_{n=1}^N\frac{(x_1+\Delta x_k-x_n^*)(z_1+\Delta z_k-z_n^*)}{d_{k,n}^4}\\
%p_{2,3}=\beta^2\sum_{k=1}^K\sum_{n=1}^N\frac{(y_1+\Delta y_k-y_n^*)(z_1+\Delta z_k-z_n^*)}{d_{k,n}^4},
%&-\frac{\beta^2}{NK}\left(\sum_{k=1}^K\sum_{n=1}^N\frac{y_1+\Delta y_k-y_n^*}{d_{k,n}^2}\right)^2\\
\qquad\qquad\qquad\quad\vdots\\
p_{3,3}=\beta^2\sum_{k=1}^K\sum_{n=1}^N\frac{(z_1+\Delta z_k-z_n^*)^2}{d_{k,n}^4}
\end{cases}
\end{equation}
and
\begin{equation}
\begin{cases}
q_{1,1}=\frac{\beta^2}{NK}\left(\sum_{k=1}^K\sum_{n=1}^N\frac{x_1+\Delta x_k-x_n^*}{d_{k,n}^2}\right)^2\\
%q_{2,2}=\frac{\beta^2}{NK}\left(\sum_{k=1}^K\sum_{n=1}^N\frac{y_1+\Delta y_k-y_n^*}{d_{k,n}^2}\right)^2\\
q_{1,2}=\frac{\beta^2}{NK}\sum_{k=1}^K\sum_{n=1}^N\frac{x_1+\Delta x_k-x_n^*}{d_{k,n}^2}\times\\
\qquad\quad\sum_{k=1}^K\sum_{n=1}^N\frac{y_1+\Delta y_k-y_n^*}{d_{k,n}^2}\\
%q_{1,3}=\frac{\beta^2}{NK}\sum_{k=1}^K\sum_{n=1}^N\frac{x_1+\Delta x_k-x_n^*}{d_{k,n}^2}\times\\
%\qquad\quad\sum_{k=1}^K\sum_{n=1}^N\frac{z_1+\Delta z_k-z_n^*}{d_{k,n}^2}\\
%q_{2,3}=\frac{\beta^2}{NK}\sum_{k=1}^K\sum_{n=1}^N\frac{y_1+\Delta y_k-y_n^*}{d_{k,n}^2}\times\\
%\qquad\quad\sum_{k=1}^K\sum_{n=1}^N\frac{z_1+\Delta z_k-z_n^*}{d_{k,n}^2}.
\qquad\qquad\qquad\quad\vdots\\
q_{3,3}=\frac{\beta^2}{NK}\left(\sum_{k=1}^K\sum_{n=1}^N\frac{z_1+\Delta z_k-z_n^*}{d_{k,n}^2}\right)^2
\end{cases}
\end{equation}

The average miss distance between the estimated position of UAV and the actual position is defined as
\begin{equation}
%\begin{aligned}
\delta=\sqrt{E[(\hat{x}_1-x_1)^2+(\hat{y}_1-y_1)^2+(\hat{z}_1-z_1)^2]}.
%&=\sqrt{E((\hat{x}_k-x_k)^2)+E((\hat{y}_k-y_k)^2)}
%\end{aligned}
\end{equation}
Since CRLB is the variance lower bound of unbiased estimator, let $\hat{\mathbf{u}}_1$ be an unbiased estimator of $\mathbf{u}_1$, i.e., $E(\hat{\mathbf{u}}_1)=\mathbf{u}_1$. According to the definition of covariance, the average miss distance satisfies the following inequality
\begin{equation}
\begin{aligned}
\delta&\geq\sqrt{{\rm CRLB}(1,1)+{\rm CRLB}(2,2)+{\rm CRLB}(3,3)}\\
&=\sigma\sqrt{\mathbf{G}^{-1}(1,1)+\mathbf{G}^{-1}(2,2)+\mathbf{G}^{-1}(3,3)}.\label{crlb}
\end{aligned}
\end{equation}
The right hand of (\ref{crlb}) is the CRLB for the average miss distance between $\hat{\mathbf{u}}_1$ and $\mathbf{u}_1$. Making use of the inverse of matrix $\mathbf{G}$, we have
\begin{equation}
\begin{aligned}
&\mathbf{G}^{-1}(1,1)=\frac{g_{2,2}g_{3,3}-g_{2,3}^2}{h},
\mathbf{G}^{-1}(2,2)=\frac{g_{1,1}g_{3,3}-g_{1,3}^2}{h},\\
&\qquad\qquad\text{and}\quad\mathbf{G}^{-1}(3,3)=\frac{g_{1,1}g_{2,2}-g_{1,2}^2}{h},
\end{aligned}
\end{equation}
where
\begin{equation}
\begin{aligned}
h=&g_{1,1}(g_{2,2}g_{3,3}-g_{2,3}^2)-g_{1,2}(g_{1,2}g_{3,3}-g_{1,3}g_{2,3})\\
&+g_{1,3}(g_{1,2}g_{2,3}-g_{1,3}g_{2,2}).
\end{aligned}
\end{equation}
Making use of the above two expressions, (\ref{crlb}) can be further reduced to the following form
\begin{equation}
\delta\geq\sigma\sqrt{\frac{g_{2,2}g_{3,3}-g_{2,3}^2+g_{1,1}g_{3,3}-g_{1,3}^2+g_{1,1}g_{2,2}-g_{1,2}^2}{h}}.
\end{equation}

\ifCLASSOPTIONcaptionsoff
  \newpage
\fi

\bibliographystyle{IEEEtran}
\bibliography{bibfile}

\end{document}